\documentclass[pdflatex,sn-mathphys-num]{sn-jnl}

\usepackage{graphicx}%
\usepackage{multirow}%
\usepackage{amsmath,amssymb,amsfonts}%
\usepackage{amsthm}%
\usepackage[title]{appendix}%
\usepackage{xcolor}%
\usepackage{textcomp}%
\usepackage{manyfoot}%
\usepackage{booktabs}%
\usepackage{algorithm}%
\usepackage{algorithmicx}%
\usepackage{algpseudocode}%
\usepackage{listings}%
\usepackage[left]{lineno}

\theoremstyle{thmstyleone}%
\theoremstyle{thmstyletwo}%

\theoremstyle{thmstylethree}%

\begin{document}

\title{Atmospheric effects on cosmic-ray muon rate at high latitude (78.9\textdegree N)}



\author[a,b]{M.~Abbrescia}
\author[c,d]{C.~Avanzini}
\author[c,d]{L.~Baldini}
\author[e,+]{R.~Baldini~Ferroli}
\author[c,d,l]{G.~Batignani}
\author[f]{M.~Battaglieri}
\author[g,h]{S.~Boi}
\author[d]{E.~Bossini}
\author[i]{F.~Carnesecchi}
\author[j]{D.~Cavazza}
\author[h]{C.~Cical\`{o}}
\author[j,k,l]{L.~Cifarelli}
\author[l]{F.~Coccetti}
\author[m]{E.~Coccia}
\author[n]{A.~Corvaglia}
\author*[o,p,l]{A.~De~Caro}\email{adecaro@unisa.it}
\author[o,p,l]{D.~De~Gruttola}
\author[o,p,l]{S.~De~Pasquale}
\author[q]{L.~Galante}
\author[l,j]{M.~Garbini}
\author[ag]{L.E.~Ghezzer}
\author[l,r]{I.~Gnesi}
\author[w]{F.~Gramegna}
\author[al]{E.~Gramstad}
\author[s,f]{S.~Grazzi}
\author[al]{E.S.~Haland}
\author[j,i]{D.~Hatzifotiadou}
\author[t,u,l]{P.~La~Rocca}
\author[l,b]{R.~Liotino}   
\author[v]{Z.~Liu}
\author[ah]{A. Lupi} 
\author[s,u]{G.~Mandaglio}
\author[j]{A.~Margotti}
\author[w]{G.~Maron}
\author[b]{M.~N.~Mazziotta}
\author[ah]{M.~Mazzola}   
\author[g,h]{A.~Mulliri}
\author[j,l]{R.~Nania}
\author[j,l]{F.~Noferini}
\author[x, ag]{F.~Nozzoli}
\author[al]{F.~Ould-Saada}
\author[k,j]{F.~Palmonari}
\author[y,n]{M.~Panareo}
\author[b]{M.~P.~Panetta}  
\author[z,d]{R.~Paoletti}
\author[aa]{C.~Pellegrino}
\author[f]{L.~Perasso}     
\author*[j,l,i]{O.~Pinazza}\email{Ombretta.Pinazza@bo.infn.it}
\author[i]{C.~Pinto}
\author[l,e]{S.~Pisano}
\author[l,e]{K.~Piscicchia}  
\author[af]{L.~Quaglia}
\author[l,t,u]{M.~Ras\`{a}}
\author[t,u,l]{F.~Riggi}
\author[ab]{G.~Righini}
\author[o,p,l]{C.~Ripoli}
\author[b]{M.~Rizzi}
\author[k,j]{B.~Sabiu}
\author[k,j]{G.~Sartorelli}
\author[j]{E.~Scapparone}
\author[ac,r]{M.~Schioppa}
\author[k,j]{G.~Scioli}
\author[z,d]{A.~Scribano}
\author[j]{M.~Selvi}
\author[aa]{A.~Shtimermann}
\author[ad,f]{M.~Taiuti}
\author[d]{G.~Terreni}
\author[s,u]{A.~Trifir\`{o}}
\author[s,u]{M.~Trimarchi}
\author[aa]{C.~Vistoli}
\author[ae]{L.~Votano}
\author[i,v]{M.~C.~S.~Williams}
\author[l,k,j,i,v,+]{A.~Zichichi}
\author[v,i]{R.~Zuyeuski}

\affil[a]{{Dipartimento di Fisica ``M. Merlin'', Universit\`a e  Politecnico di Bari, Via Amendola 173, 70125 Bari, Italy}}

\affil[b]{{INFN, Sezione di Bari, Via Orabona 4, 70126 Bari, Italy}}

\affil[c]{{Dipartimento di Fisica ``E. Fermi'', Universit\`a di Pisa, Largo Bruno Pontecorvo 3, 56127 Pisa, Italy}}

\affil[d]{{INFN, Sezione di Pisa, Largo Bruno Pontecorvo 3, 56127 Pisa, Italy}}

\affil[e]{{INFN, Laboratori Nazionali di Frascati, Via Enrico Fermi 54, 00044 Frascati (RM), Italy}}

\affil[f]{{INFN, Sezione di Genova, Via Dodecaneso 33, 16146 Genova, Italy}}

\affil[g]{{Dipartimento di Fisica, Universit\`a di Cagliari, S.P. Monserrato-Sestu km 0.700, 09042 Monserrato (CA), Italy}}

\affil[h]{{INFN, Sezione di Cagliari, S.P. Monserrato-Sestu Km 0.700, 09042 Monserrato (CA), Italy}}

\affil[i]{{European Organisation for Nuclear Research (CERN), Esplanade des Particules 1, 1211 Geneva 23, Switzerland}}

\affil[j]{{INFN Sezione di Bologna, Viale Carlo Berti Pichat 6/2, 40127 Bologna, Italy}}

\affil[k]{{Dipartimento di Fisica e Astronomia ``A. Righi'', Universit\`a di Bologna, Viale Carlo Berti Pichat 6/2, 40127 Bologna, Italy}}

\affil[l]{{Museo Storico della Fisica e Centro Studi e Ricerche ``E. Fermi'', Via Panisperna 89a, 00184 Roma, Italy}}

\affil[m]{{Gran Sasso Science Institute, Viale Francesco Crispi 7,  67100 L'Aquila, Italy}}

\affil[n]{{INFN Sezione di Lecce, Via per Arnesano, 73100 Lecce, Italy}}

\affil[o]{{Dipartimento di Fisica ``E. R. Caianiello'', Universit\`a di Salerno, Via Giovanni Paolo II 132, 84084 Fisciano (SA), Italy}}

\affil[p]{{INFN Gruppo Collegato di Salerno, Via Giovanni Paolo II 132, 84084 Fisciano (SA), Italy}}

\affil[q]{{Teaching and Language Lab, Politecnico di Torino, Corso Duca degli Abruzzi 24, Torino, Italy}}

\affil[r]{{INFN Gruppo Collegato di Cosenza, Via Pietro Bucci, Rende (Cosenza), Italy}}

\affil[s]{{Dipartimento di Scienze Matematiche e Informatiche, Scienze Fisiche e Scienze della Terra, Universit\`a di Messina, Viale Ferdinando Stagno d'Alcontres 31, 98166 Messina, Italy}}

\affil[t]{{Dipartimento di Fisica e Astronomia ``E. Majorana'', Universit\`a di Catania, Via S. Sofia 64, 95123 Catania, Italy}}

\affil[u]{{INFN Sezione di Catania, Via S. Sofia 64, 95123 Catania, Italy}}

\affil[v]{{ICSC World Laboratory, Geneva, Switzerland}}

\affil[w]{{INFN Laboratori Nazionali di Legnaro, Viale dell'Universit\`a 2, 35020 Legnaro, Italy}}

\affil[x]{{INFN Trento Institute for Fundamental Physics and Applications, Via Sommarive 14, 38123 Trento, Italy}}

\affil[y]{{Dipartimento di Matematica e Fisica ``E. De Giorgi'', Universit\`a del Salento, Via per Arnesano, 73100 Lecce, Italy}}

\affil[z]{{Dipartimento di Scienze Fisiche, della Terra e dell'Ambiente, Universit\`a di Siena, Via Roma 56, 53100 Siena, Italy}}

\affil[aa]{{INFN-CNAF, Viale Carlo Berti Pichat 6/2, 40127 Bologna, Italy}}

\affil[ab]{{CNR, Istituto di Fisica Applicata ``Nello Carrara'', Via Madonna del Piano 10, 50019 Sesto Fiorentino (FI), Italy}}

\affil[ac]{{Dipartimento di Fisica, Universit\`a della Calabria, Via Pietro Bucci, Rende (CS), Italy}}

\affil[ad]{{Dipartimento di Fisica, Universit\`a di Genova, Via Dodecaneso 33, 16146 Genova, Italy}}

\affil[ae]{{INFN Laboratori Nazionali del Gran Sasso, Via G. Acitelli 22, 67100 Assergi (AQ), Italy}}

\affil[al]{{Physics Department, Oslo University, P.O.Box 1048, 0316 Oslo, Norway}}

\affil[af]{{INFN Sezione di Torino, Via Pietro Giuria 1, 10125 Torino, Italy}}

\affil[ag]{{Dipartimento di Fisica Universit\`a di Trento, via Sommarive 14, 38123 Trento, Italy}}

\affil[ah]{{National Research Council, Institute of Polar Sciences, Bologna, 40129, Italy}}

\affil[+]{{Deceased}}
%
%
%
%
%


\abstract{
Since 2019, three scintillator detectors of the EEE collaboration have been continuously measuring cosmic muon rates at 78.9°N at the Ny-Ålesund Research Station (Svalbard). The resulting six-year time series reveals a pronounced annual modulation, driven primarily by seasonal atmospheric variations. Utilizing routine radiosonde profiles collected above the same site, we applied several established techniques —along with a tailored analysis approach— to investigate the relationship between muon rate and atmospheric temperature. 
The temperature-corrected muon-rates are analysed using the Lomb-Scargle periodogram technique in order to investigate the presence of remaining periodic structures. Finally, the temperature corrections coefficients of our analysis are compared with measurements in other stations located at lower latitudes.
}


\maketitle

\section*{Dedication}
This article is dedicated to the memory of Professor Antonino Zichichi (1929--2026), whose visionary ideas made this work possible.

\section{Introduction}\label{intro}

For more than twenty years, the Extreme Energy Events (EEE) Collaboration has been carrying out scientific research projects with a strong educational component \cite{AZ}. These activities are supported by the Enrico Fermi Historical Museum of Physics and Research and Study Centre (CREF) and by the Italian National Institute for Nuclear Physics (INFN).

Over this period, approximately fifty muon telescopes based on gaseous Multigap Resistive Plate Chambers (MRPCs) have been designed, assembled, and operated with the active involvement of teachers and students from Italian high schools \cite{EEE20}. These detectors are mainly installed inside the schools and are distributed across the Italian territory. 

In 2018, the EEE Collaboration developed a set of scintillator-based transportable detectors, named POLA-R, which were used for a series of measurements of the cosmic-ray flux at different latitudes \cite{POLAR, POLARlatitude}. Since 2019, three POLA-R units were deployed at the Ny-Ålesund Research Station\footnote{{Ny-Ålesund Research Station, https://nyalesundresearch.no/}}  (Svalbard) and have been continuously recording the muon flux at 78.9\textdegree N.

For the detectors at Ny-Ålesund, the pressure-corrected time series accumulated over six years exhibit a clear annual periodicity \cite{POLAperiodicity}, primarily driven by variations in atmospheric parameters, as extensively documented in the literature \cite{Dorman, deMendonça_2016}. In this work, we investigate in detail the impact of atmospheric temperature on the muon rate, exploiting daily atmospheric profiles above Ny-Ålesund measured by the Alfred Wegener Institute for Polar and Marine Research (AWI) using radiosondes \cite{maturilli2025hrrm}.

\section{POLA-R detectors, data processing and corrections}

\subsection{POLA-R detectors installed at Ny-Ålesund}

The POLA-R detectors consist of two scintillator planes (40x60 cm$^2$), each subdivided into four equal tiles. Signals from each tile are read out by two silicon photo-multipliers (SiPMs). These detectors provide limited tracking capability but achieve a time resolution of tens of nanoseconds, exploited when defining the coincidences between correspondent tiles. A detailed description of their design and performance is available in \cite{POLAR, POLARlatitude, EEEsvalbard, POLAperiodicity} together with a few physics results.

Three detectors —POLA-1, POLA-3 and POLA-4— are installed at three sites, managed by the Italian National Research Council (CNR) within the Ny-Ålesund Research Station, Svalbard. They are separated by distances ranging from 500 to 1000 meters and are located only a few dozen meters above sea level. Each detector is equipped with a GPS and multiple sensors for environmental measurements.

\subsection{Data acquisition and storage}

Muon data from the POLA-R detectors are initially stored locally, then transferred to the INFN-CNAF computing center in Bologna, Italy. Here, data are calibrated, corrected, and binned into 1-minute intervals, along with GPS timestamps and environmental parameters. Each event is assigned a quality flag to identify potential issues like incorrect sensor readings or abnormal integrated signal durations.

\subsection{Event selection, efficiency and barometric corrections}
\label{sect:cor-pressure}

Only high-quality events are used in the analysis. Events with faulty pressure readings or high temperatures at sensor level —indicative of malfunctioning electronics — are excluded. The detectors use a majority trigger logic (3 out of 4 SiPMs must detect a signal, requesting signals on both planes). A global efficiency correction is applied based on measured SiPM performance (see \cite{POLAperiodicity} for details).
No corrections for acceptance are applied, since only relative variations over time are studied.

To account for the well-known negative correlation between atmospheric pressure and the muon rate —arising from the exponential decrease of the pressure with altitude— a correction can be applied using surface atmospheric pressure measurements. The barometric correction coefficient is obtained via linear regression between deviations in the muon rate and the logarithm of the ratio between the pressure and its average. This coefficient is calculated individually for each detector over the entire dataset. Further details on the barometric correction procedure can be found in \cite{POLAperiodicity}.

\begin{figure}[hbt!]
    \centering
    \includegraphics[width=1\linewidth]{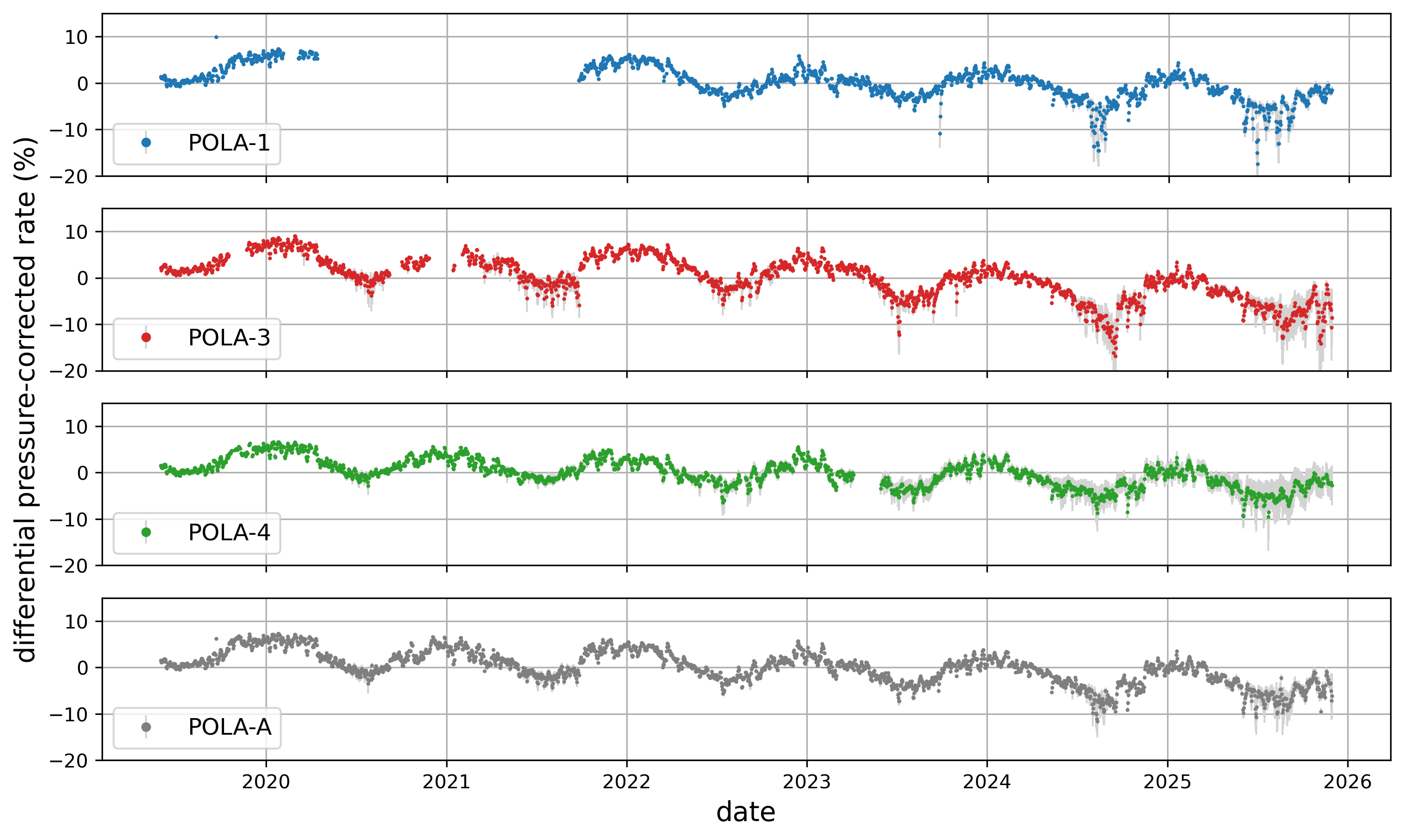}
    \caption{Differential rates of detectors POLA-1, POLA-3 and POLA-4, corrected for atmospheric pressure. The lower plot shows the average rate of the three, indicated as POLA-A. }
    \label{fig:rates}
\end{figure}

Figure~\ref{fig:rates} shows the differential, pressure-corrected muon rate time series at 1-day resolution for the three detectors and their combined average POLA-A. The \textit{differential rate} is defined as the relative deviation from the mean rate $I$, computed as $100 \times (I-\langle I\rangle)/\langle I\rangle$, and is expressed in percent. The total error combines statistical errors and systematic errors as calculated in reference \cite{POLAperiodicity}.

The three detectors exhibit fully consistent behavior; therefore, for this study we used only the mean value. During periods when data from one or more detectors are unavailable, the mean is computed using only the available data, and the uncertainty is evaluated accordingly.

\section{Atmospheric effects on muons}
\label{sect:athmosphericeffects}

The dependence of ground-level muon flux on atmospheric conditions was first recognized nearly a century ago, and since then numerous measurements and models have been developed to account for this effect in cosmic-ray rate analyses.
Usually, atmospheric pressure and temperature corrections are treated separately, although strictly connected \cite{Dorman}. 
A comprehensive review of the corresponding literature can be found in  \cite{deMendonça_2016}. A discussion of the dynamics of the atmospheric levels mainly involved in the muon production may be found in \cite{Tanaka}.

In this context, for POLA-R data collected in Ny-Ålesund, after applying the pressure correction described in Section \ref{sect:cor-pressure} and in \cite{POLAperiodicity}, the effect of temperature variations on muon flux at different altitudes must be evaluated.

In general, the correlation between temperature and muon rate is considered to arise from both negative and positive contributions. The negative contribution occurs because atmospheric expansion at higher temperatures shifts muon production to higher altitudes, increasing the average decay path and decay probability —particularly for low-energy muons— thereby reducing the observed muon rate. 

The positive contribution arises from the reduced atmospheric density, which favors pion decay against pion re-interaction, increasing the rate of high-energy muons: this effect is thus especially relevant for underground experiments \cite{LVD19}.

In the case of the POLA-R detectors, which are installed at a site with near-zero geomagnetic cutoff —where low-energy muons are predominant— the effect of temperature variations should be essentially negative. 
Moreover, the structure of the atmosphere at polar latitudes is highly peculiar, making the study of the temperature effect on the POLA-R rates particularly interesting.

\subsection{The atmospheric temperature above Ny-Ålesund}

\begin{figure}[hbt!]
    \centering
    \includegraphics[width=1\linewidth]{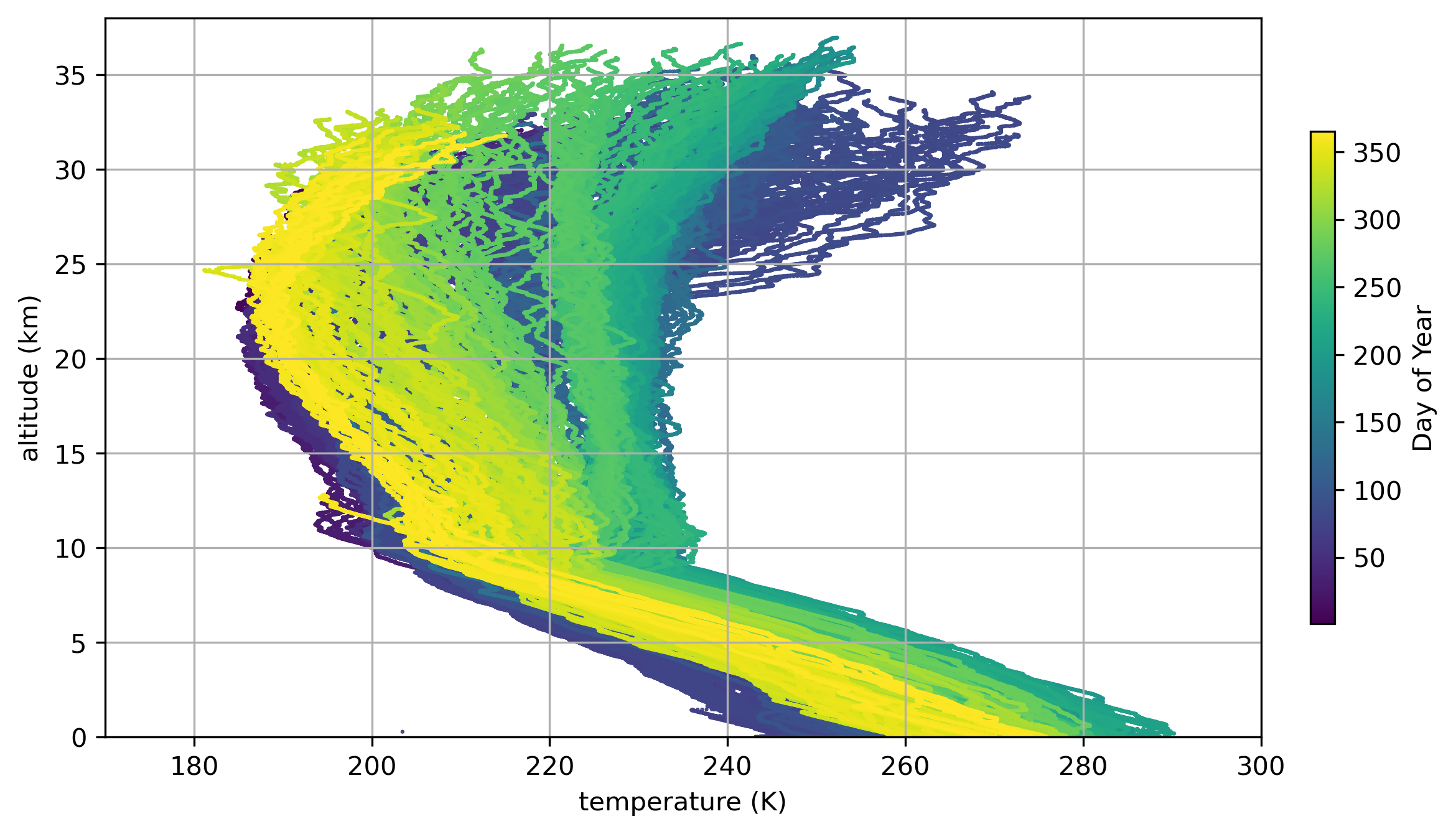}
    \caption{The temperature profile above Ny-Ålesund during 2020. Different colours identify days of the year, from January to December 2020. Data taken from \cite{maturilli2025hrrm}.}
    \label{fig:atmospheric_temperature_profile}
\end{figure}

The AWI has been conducting a long-term, regular radiosonde (RS) monitoring for several decades to monitor the atmospheric profile above Ny-Ålesund. This research town, located on Spitsbergen Island within the Svalbard archipelago (78.9\textdegree N, 11 m a.s.l.), serves as a crucial monitoring site for the transition from mid-latitudes to the polar region. Measurements of air temperature and relative humidity were routinely performed—usually once per day at 12:00 UTC—using Vaisala RS90 and RS92 radiosondes. These sensors provide observational data up to altitudes as high as 31-38 km, allowing for the derivation of atmospheric parameter profiles as a function of altitude or pressure levels with high vertical resolution \cite{Maturilli-paper}.
The vertical profiles of temperature obtained during 2020 \cite{maturilli2025hrrm} are illustrated in Figure \ref{fig:atmospheric_temperature_profile}. These measurements reveal substantial seasonal variability in the atmospheric conditions above the site. Surface-level air temperature (T0) undergoes clear seasonal cycles, ranging approximately from a minimum of -10.8 K in March to a maximum of +7.1 K in July. Notably, the climate conditions at Ny-Ålesund are partially mitigated by the warm effects of the North Atlantic Drift, and T0 exhibited a clear warming trend of approximately +1.3 K per decade between 1994 and 2013 \cite{Maturilli-paper}.
In the middle and upper atmosphere, the tropopause temperature minimum (T$_{min}$) varies between approximately 200 K during winter and 230 K during summer, with the height of this minimum (z$_{min}$) typically ranging from 8 km in spring to 9 km in summer. Crucially, the thermal inversions observed near the ground in Ny-Ålesund during winter and spring are generally weaker compared to those monitored at continental Arctic sites. As shown in Figure \ref{fig:atmospheric_temperature_profile}, the stratospheric region above 20 km exhibits the most dramatic thermal excursions, with variations up to 50 K observed during the year. The availability of these comprehensive vertical profiles is essential for investigating different models designed to correct muon rate measurements, as they cover the full atmospheric volume above the POLA-R detectors.

\section{Atmospheric correction models based on temperature or altitude }
\label{sect:LinMod}

Using temperature trend data together with the associated atmospheric parameters over the entire column above the site makes it possible to assess different correction models for muon rate measurements. In literature different approaches have been proposed: the linear empirical model methods, where the correction is based on a unique parameter at fixed altitude or pressure, and the integral models, where the entire profile in altitude is considered. A summary and a more detailed description of these models may be found in \cite{deMendonça_2016}. In the following two examples for each case will be studied. 

\subsection{Linear empirical models}

Models for the linear correction of the muon rate separate the pressure and atmospheric temperature effect and have this form:

\begin{equation}
    \Delta I_{PTC} = \Delta I_{PC} -\Delta I_T
        \label{eq:ratecorre}
\end{equation}
where  $\Delta I_{PTC}$ is the muon rate variation corrected for pressure $P$ and
temperature $T$, $\Delta I_{PC}$ the rate variation corrected for pressure only, and 
\begin{equation}
    \Delta I_T = \alpha_T *\Delta T
\end{equation}
is the rate variation due to the atmospheric temperature $T$ variations.
The determination of $\alpha_T$  is done measuring for each model the correlation
\begin{equation}
    \Delta I_{PC} = \alpha_{model} *\Delta T_{model}
\end{equation}

\begin{figure} [hbt!]
    \centering
    \includegraphics[width=1\linewidth]{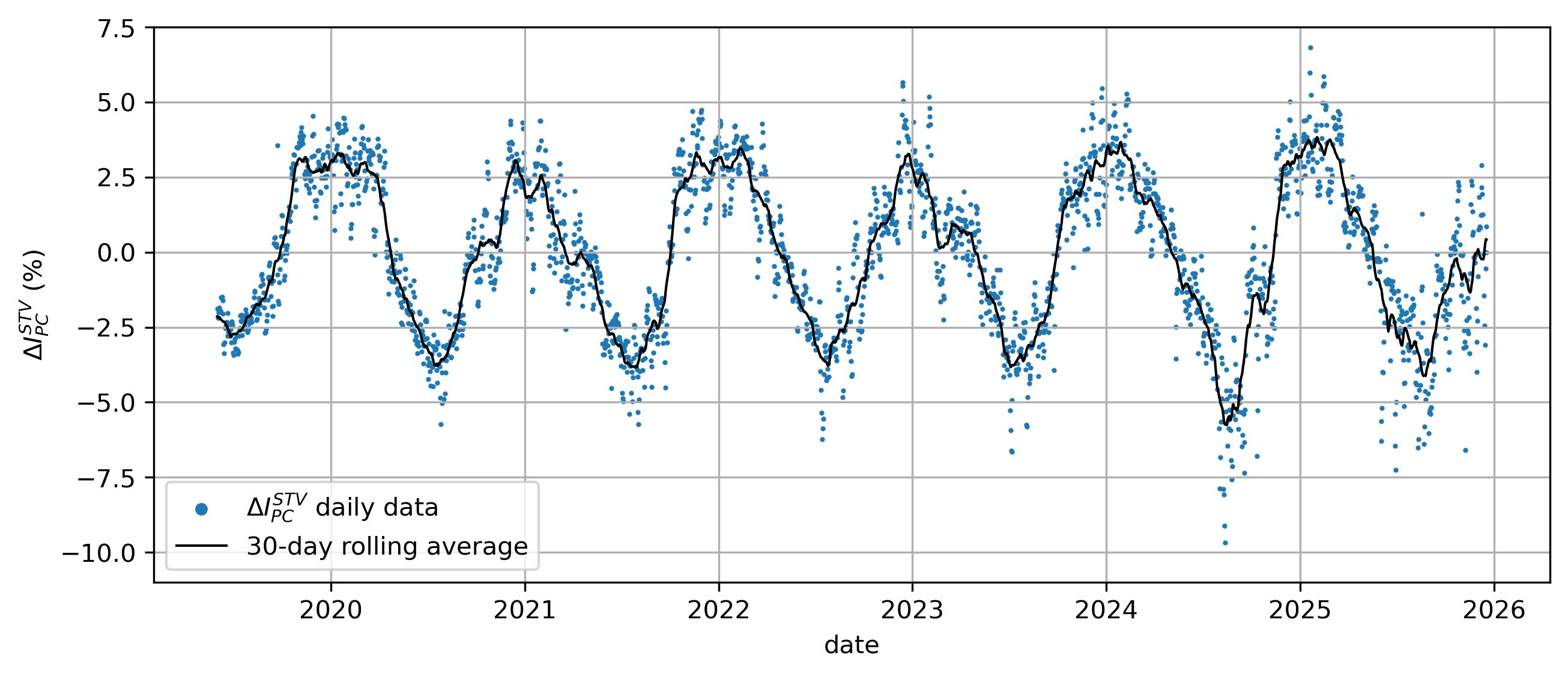}
    \caption{The short-time variation differential rate as measured with the POLA-R detectors (derived from results of \cite{POLAperiodicity}). To highlight the underlying trend of the measured points, the solid black curve visualizes a 30-day rolling average of the daily data.}
    \label{fig:Drate_src}
\end{figure}

To remove the modulation associated with the solar cycle, rather than considering $\Delta I_{PC}$ with respect to the average over the full six-year period, we adopt the approach described in \cite{deMendonça_2016}, in which a rolling average over a relatively long period is subtracted from the muon rates.
In the present analysis,  a 24-month rolling average around the time $t$, $I_{roll}(t)$, was found to effectively suppress the solar-cycle variations while avoiding the introduction of spurious oscillations in the data. The border effects at the beginning and end of the six-year period do not affect the final result in a  noticeable way. 
Using the reference data from \cite{POLAperiodicity} for $I_{PC}$, the result of this subtraction is shown in 
Figure \ref{fig:Drate_src}, which isolate the short-term yearly variations of the pressure-corrected rate, $\Delta I_{PC}^{STV}$, with a time resolution of one day, defined as:

\begin{equation}
    \Delta I_{PC}^{STV}(t) = 100*\displaystyle\frac{I_{PC}(t) - I_{roll}(t)}{I_{roll}(t)}
\label{eq:dratec_srt}
\end{equation}

The value  $\alpha_{model}$ is evaluated from the slope of the linear regression between the short time variation of the pressure-corrected rate $\Delta I_{PC}^{STV}$ and the differential atmospheric temperature parameter used.

\subsubsection{The Atmospheric Expansion (Duperier) method -- ATE} 

\begin{figure} [hbt!]
    \centering
    \includegraphics[width=1\linewidth]{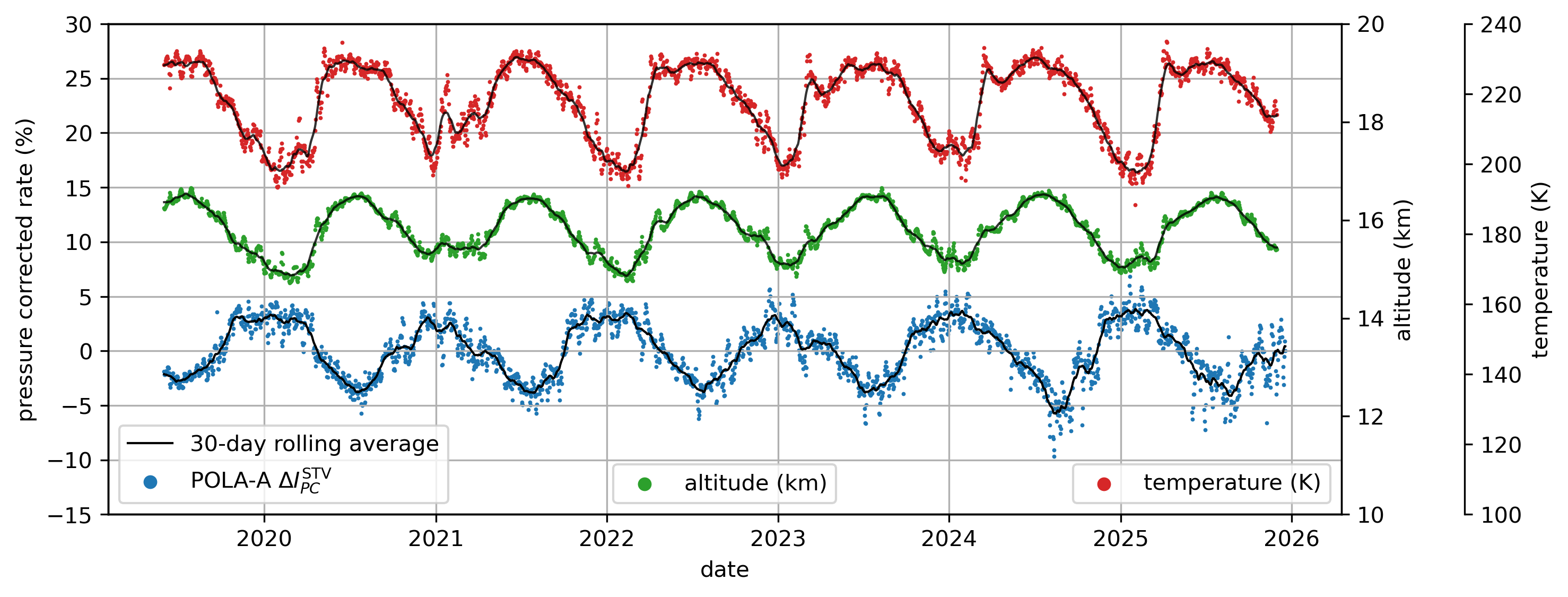}
    \caption{Altitude of the 100 hPa isobar (green) and the corresponding temperature (red). In blue, the differential rate for short-term variations. All 1-day time series are plotted together with the 30-day rolling average (black lines), shown to highlight the underlying trend.}
    \label{fig:isobare_100hPa}
\end{figure}

This model was first proposed by A. Duperier in 1949 \cite{Duperier_1949}.
In this model, the atmospheric temperature is the one corresponding to the altitude of the isobar at 100 hPa, which is considered the level where the maximum muon production occurs. In the literature, it is customary to use, as a parameter, not the temperature itself but rather the altitude of the corresponding isobar.
According to the AWI data, the trend of this altitude is shown in Figure \ref{fig:isobare_100hPa} together with the corresponding temperature trend. The anti-correlation of the muon rate with both temperature or pressure is evident.

\begin{figure} [hbt!]
    \centering
    \includegraphics[width=1\linewidth]{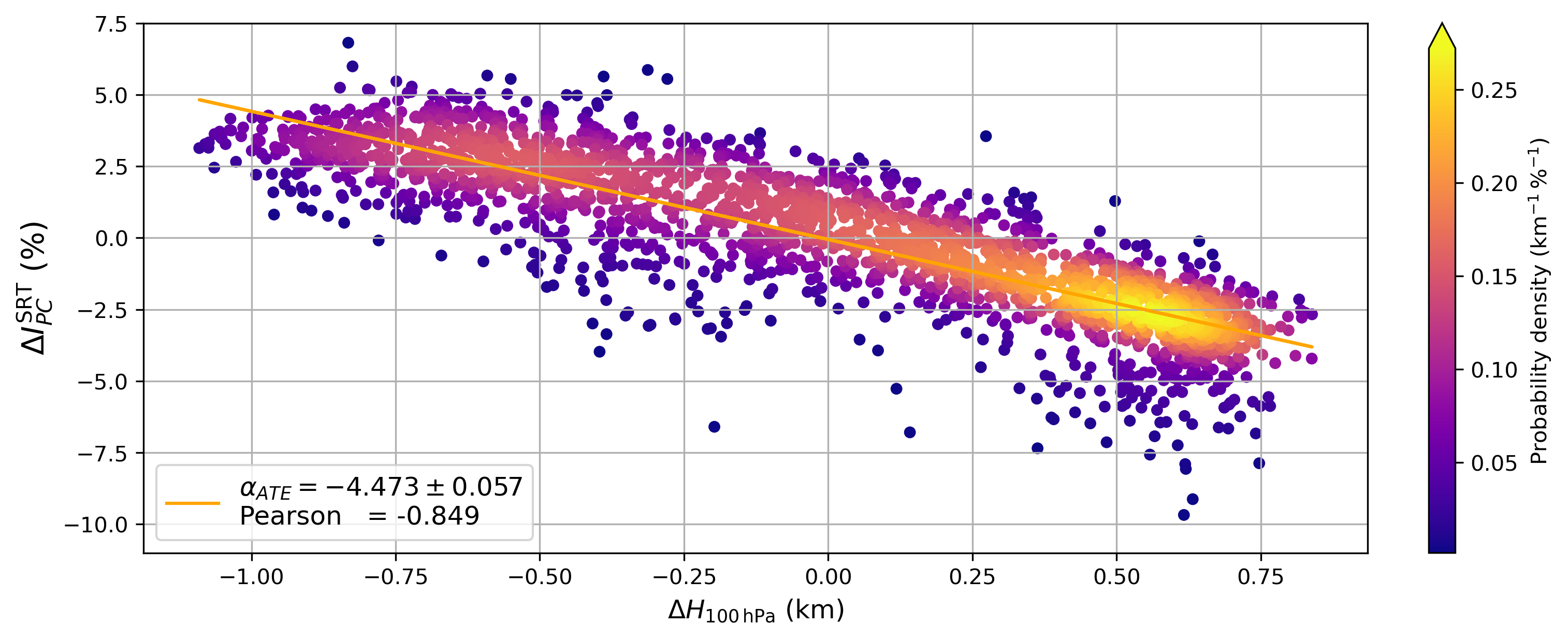}
    \caption{Correlation between $\Delta I_{PC}^{STV}$ and differential altitude of the 100 hPa isobar, from which the value of correction coefficient $\alpha_{ATE}$ is derived.}
    \label{fig:alpha_ate}
\end{figure}

In this case, $\alpha$ is the slope of the linear regression between the short-time variation of the pressure-corrected rate $\Delta I_{PC}^{STV}$ (eq.~\eqref{eq:dratec_srt}) and the variation of the altitude of the 100 hPa isobar, $\Delta H_{100 hPa}$, according to 

\begin{equation}
\begin{aligned}
\Delta I_{PC}^{\mathrm{STV}} &= \alpha_{ATE}\,\Delta H_{100\,\mathrm{hPa}}, \\
\text{where}\quad
\Delta H_{100\,\mathrm{hPa}} &= H_{100\,\mathrm{hPa}} - \langle H_{100\,\mathrm{hPa}} \rangle .
\end{aligned}
\end{equation}
where $\alpha_{ATE}$ units are $\%/km$. The result is reported in  Figure \ref{fig:alpha_ate}.

Note that in Figure \ref{fig:isobare_100hPa} the  100 hPa altitude trend is somehow smoother than the temperature one, which shows more peculiar structures like those at the beginning of each year. We have verified that, computing the correction as a function of the temperature at 100 hPa isobar instead of the altitude, produces results a bit different and very similar to the ones obtained in the Maximum Muon Production method, described in the next Section.

\subsubsection{The Maximum Muon Production method -- MMP}

\begin{figure} [hbt!]
    \centering
    \includegraphics[width=1\linewidth]{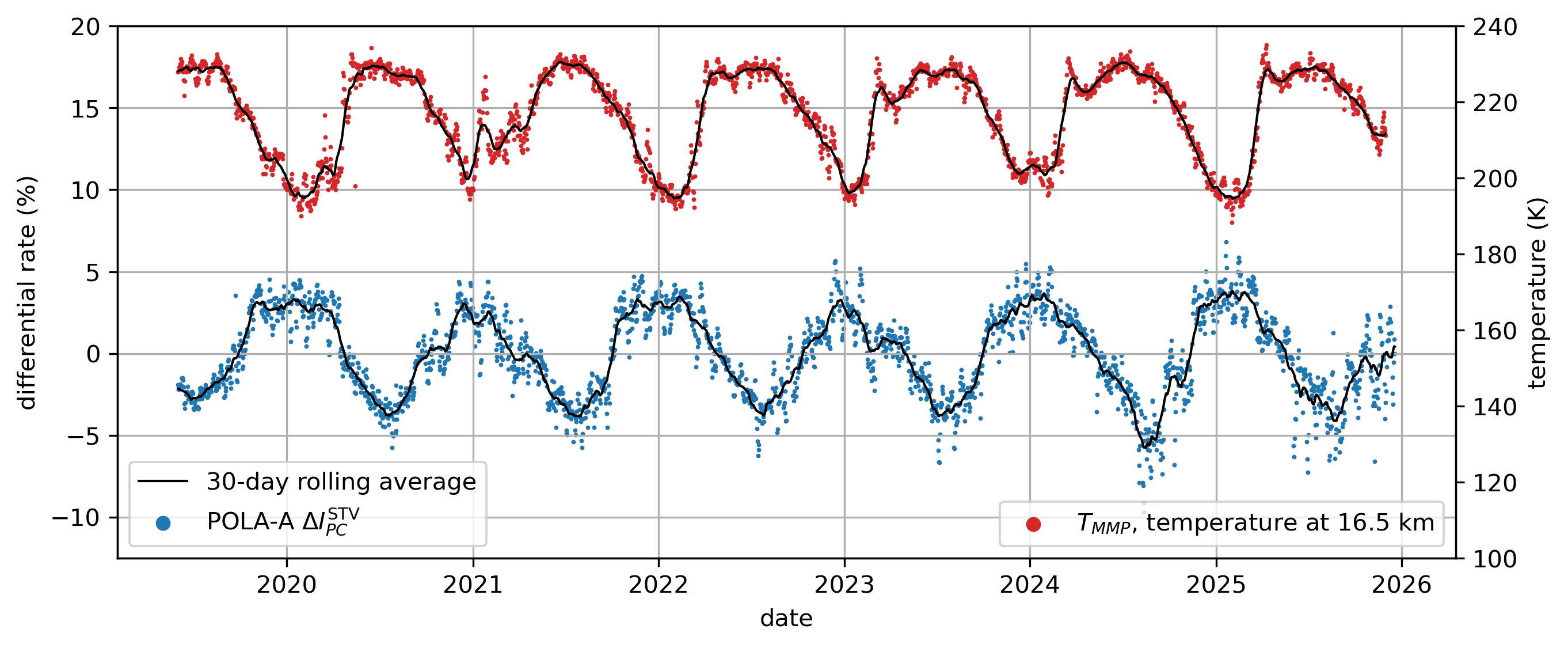}
    \caption{In red, the temperature profile at 16.5 km altitude (used by the MMP method). In blue, the differential rate for short-term variations. All 1-day time series are plotted together with the 30-day rolling average (black lines), shown to highlight the underlying trend.}
    \label{fig:100hpa-16500m}
\end{figure}

This method is based on the temperature deviation observed at the MMP altitude, $h_{MMP}$, where the maximum of the cosmic-ray shower development is assumed to occur.  Considering $h_{MMP}$ to be 16.5 km, the corresponding temperature is reported in Figure \ref{fig:100hpa-16500m} and also in this case a clear anti-correlation is observed. The correlation coefficient is measured using the expression 

\begin{equation}
\Delta I_{PC}^{STV} = \alpha_{MMP}*\Delta T[h_{MMP}] 
\end{equation}
where $\alpha_{MMP}$ units are $\%/K$.

\begin{figure} [hbt!]
    \centering
    \includegraphics[width=1\linewidth]{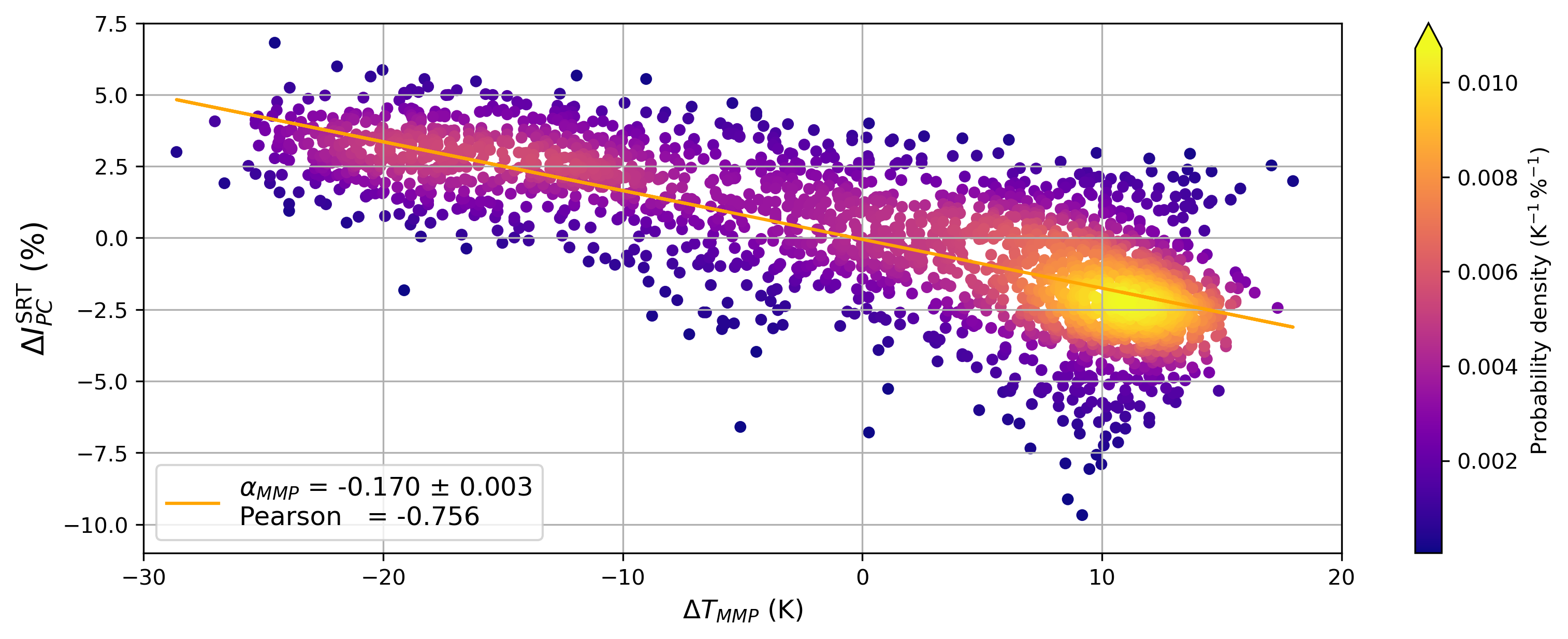}
    \caption{Correlation between $\Delta I_{PC}^{STV}$ and the differential temperature at 16.5 km altitude, from which the correction coefficient $\alpha_{MMP}$ is derived. }
    \label{fig:alpha_MMP}
\end{figure}

The result is shown in Figure \ref{fig:alpha_MMP} where a negative coefficient is measured. 
Using a fixed altitude like in this method does not account for local atmospheric characteristics that can influence the altitude of maximum cosmic-ray shower development. For example, the tropopause altitude is approximately 17 km at the equator and about 10 km in polar regions. Fixing $h_{MMP}$ to be 16.5 km may modify, at changing latitude, the balance between the negative and positive contributions described in Section \ref{sect:athmosphericeffects}. 

\subsection{Integral and discrete methods}

These methods are not based on a single measured parameter (pressure or altitude as in the previous methods), but take into account the entire vertical profile, with a weight coefficient depending on the mass, the pressure or the altitude level.

\subsubsection{The Mass Weighted method -- MSS}

\begin{figure} [hbt!]
    \centering
    \includegraphics[width=1\linewidth]{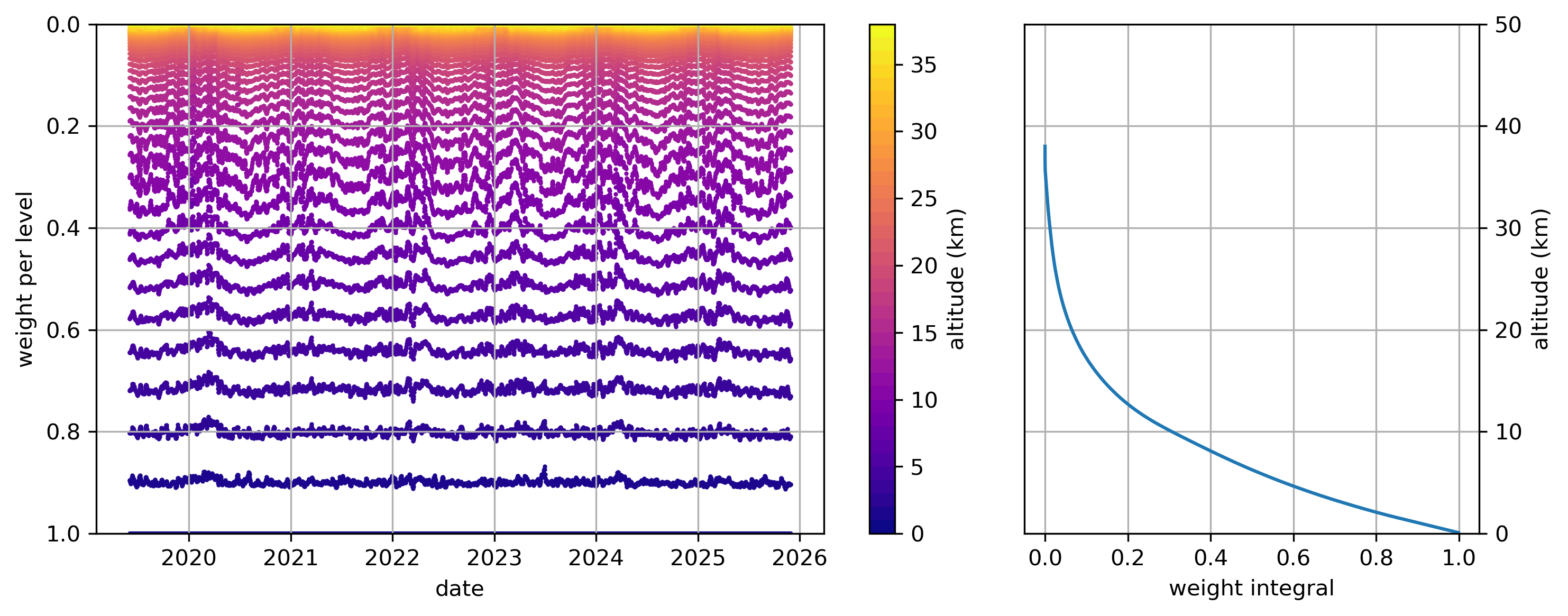}
    \caption{ MSS method. On the left side, the weight functions $w[h_i]$ per level, displayed every 1000 m altitude, as a function of time (6 years). On the right side, the integral of weight functions, normalized to the maximum value. For radiosonde files we have data up to 38 km.}
    \label{fig:w_i}
\end{figure}

In this method \cite{MSS-1,MSS-2} the muon intensity variation due to temperature effects is defined as:
\begin{equation}
   \Delta I_{PC}^{STV} = \alpha_{MSS} * \Delta T_{MSS}
\end{equation}
where $\alpha_{MSS}$ is in $\%/K$ and 
\begin{equation}
\label{eq:discrete}
 T_{MSS} = \displaystyle\sum_{i=0}^n w[h_i] *T[h_i] ,  
\end{equation}
with $h_i$ the altitude bin and  $w[h_i] = \displaystyle\frac{x[h_i] - x[h_{i+1}]}{x[h_0]}$  (being $x[h_i]$ the atmospheric depth at altitude $h_i$ in units  g/cm$^2$). $w[h_i]$ is the air mass weighted function  normalized to  $[0,1]$ range and graphically reported in Figure \ref{fig:w_i}. Note the relatively constant values of the weight per level over time, which reflect in a smooth weight integral, as shown in the right panel of Figure \ref{fig:w_i}.
\begin{figure} [hbt!]
    \centering
    \includegraphics[width=1\linewidth]{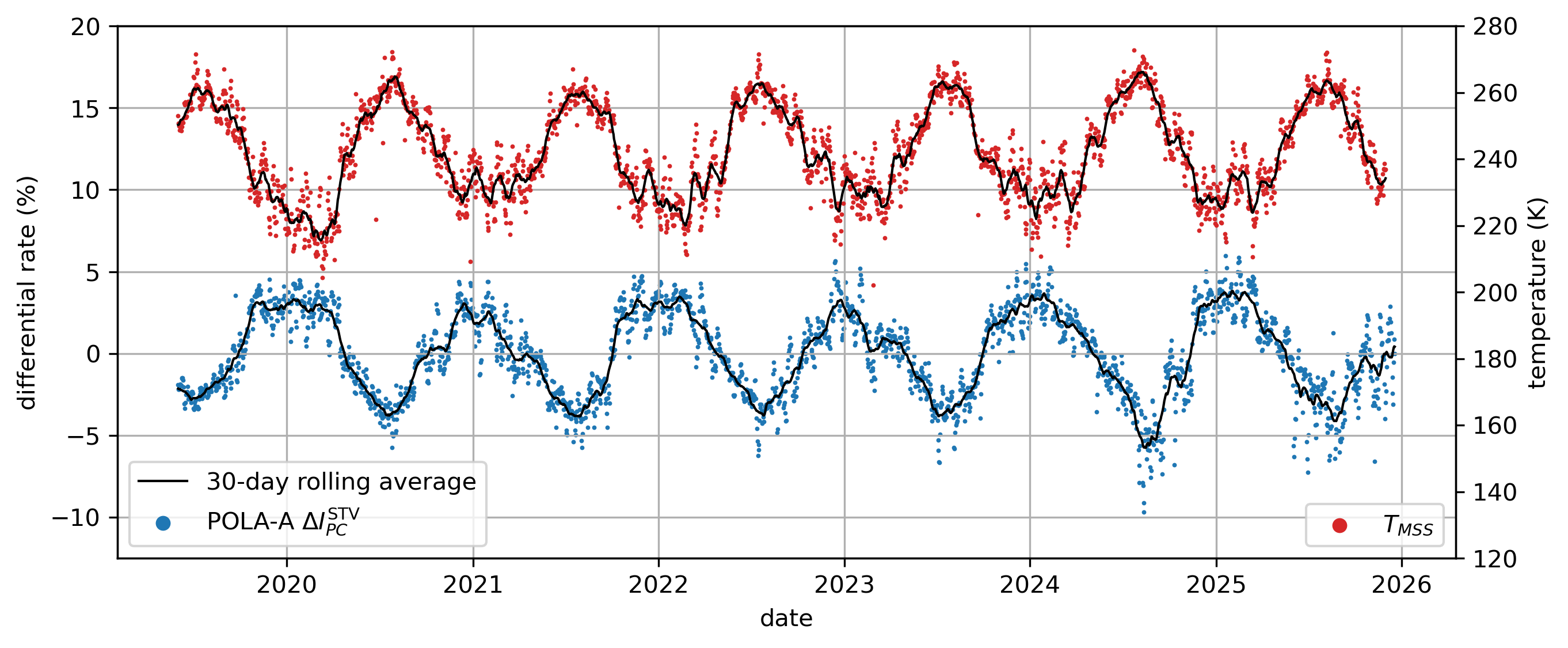}
    \caption{The temperature $T_{MSS}$ (red), obtained by weighting the different layers of the atmospheric profile. In blue, the differential rate for short-term variations. All 1-day time series are plotted together with their 30-day rolling average (black lines), shown to highlight the underlying trend.}
    \label{fig:tempMSS}
\end{figure}

\begin{figure} [hbt!]
    \centering
    \includegraphics[width=1\linewidth]{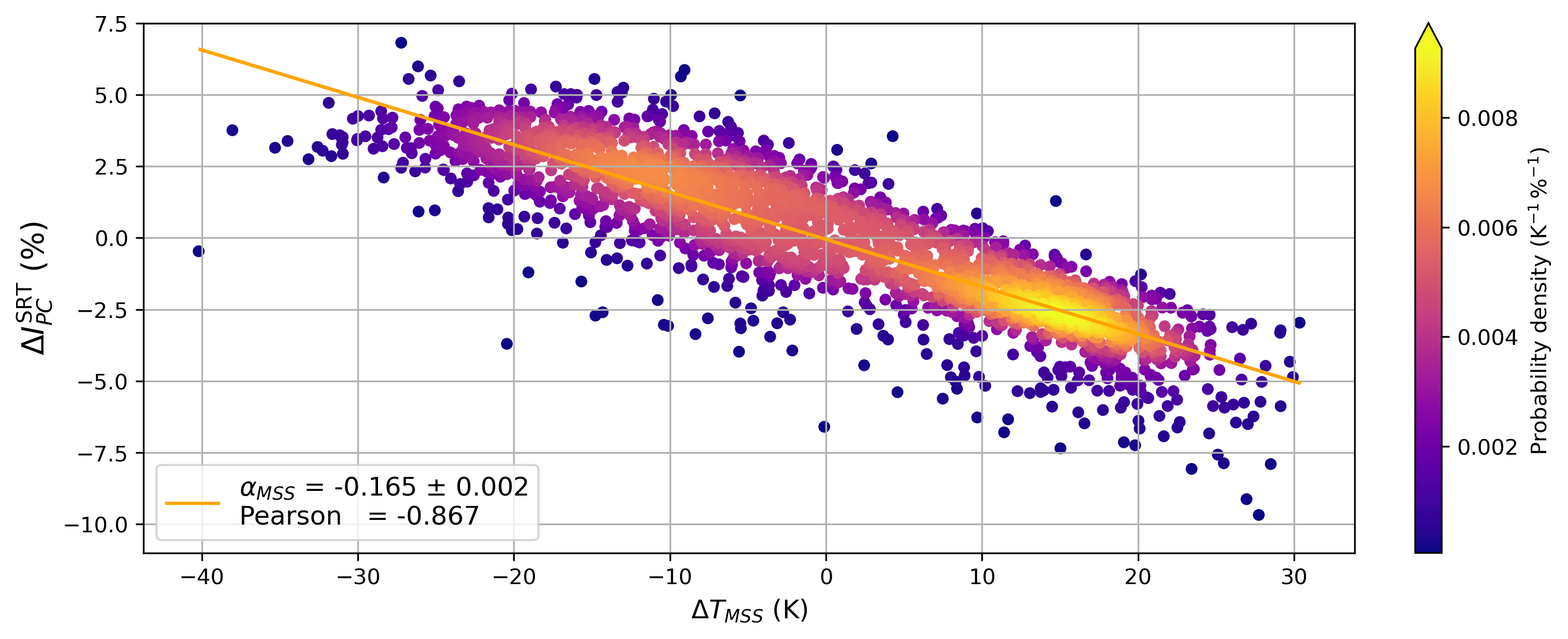}
    \caption{Correlation between $\Delta I_{PC}^{STV}$ and differential temperature $T_{MSS}$.}
    \label{fig:alpha_MSS}
\end{figure}

The anti-correlation between the differential rate and the temperature $T_{MSS}$ is clearly observed in Figure  \ref{fig:tempMSS}. 

The $\alpha_{MSS}$ coefficient is computed fitting the data reported in Figure \ref{fig:alpha_MSS}.

In this model the weight depends on the altitude and air densities (therefore mainly on the pressure at each layer), but does not take into account  other specific feature of the atmospheric structure (latitude dependent) above the detectors.

\subsubsection{The  Discrete Correlation Method -- DCM }

To account for the local characteristics of the atmosphere, we propose a new empirical model, which we call the Discrete Correlation Method (DCM), based on the correlation of data with temperatures in layers at different pressure levels. In our model, the weight used to calculate the reference temperature (see equation \eqref{eq:discrete}) comes from the Pearson correlation coefficient, suitably normalized. 

We recall that the Pearson correlation coefficient measures the linear correlation between two series $X$ and $Y$ and is defined as $\rho_{X,Y} = {\operatorname{Cov}(X,Y)}/{\sigma_X \sigma_Y}$.
It takes values in $[-1,1]$: positive values indicate a positive linear correlation, negative values indicate anti-correlation and values close to zero indicate weak or no linear correlation.

\begin{figure}
    \centering
    \includegraphics[width=1\linewidth]{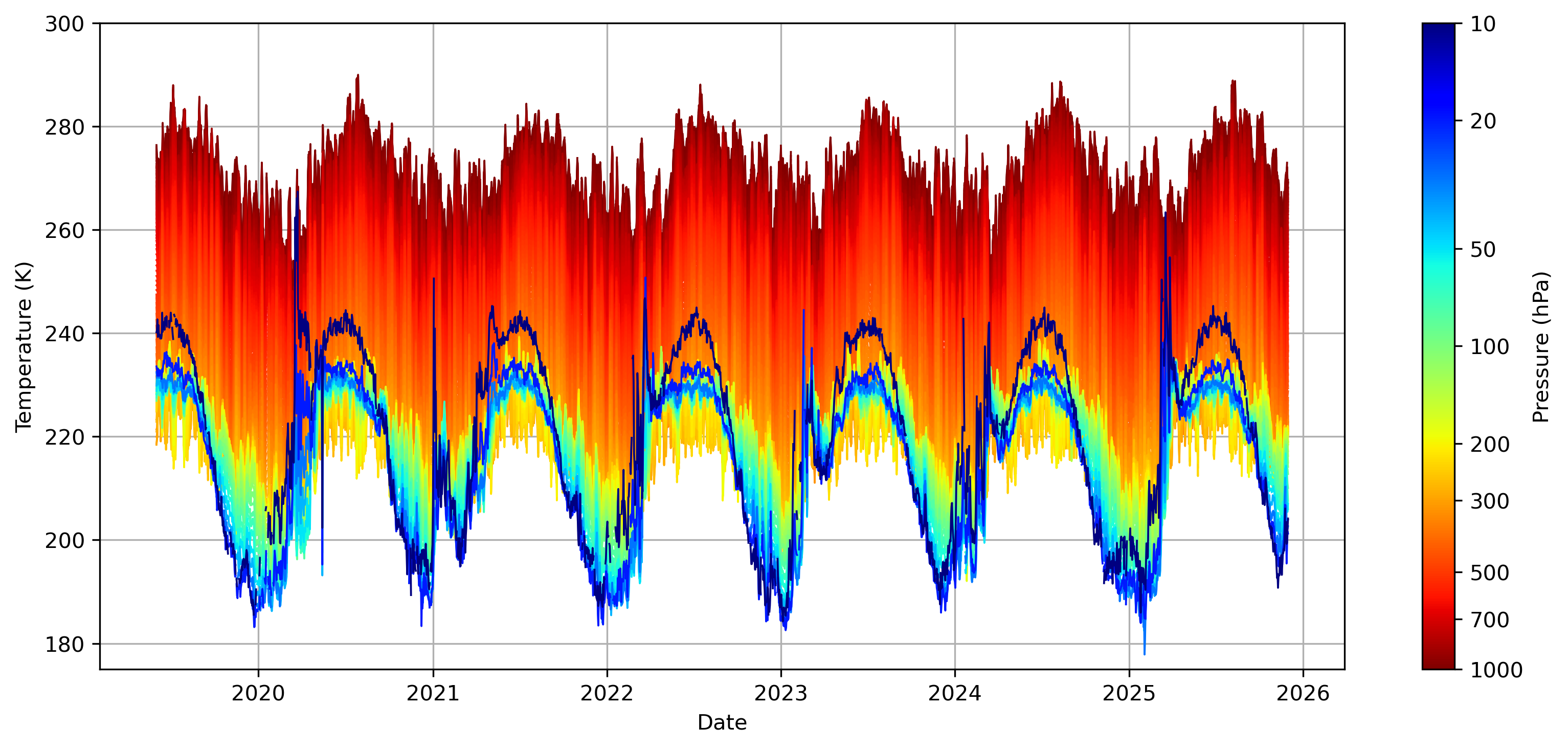}
    \caption{The temperature trends at \texttt{\textasciitilde}100 pressure levels, one every 10 hPa, from data recorded by radiosonde at 78.9°N above Ny-Ålesund.}
    \label{fig:DCM temperature pressure}
\end{figure}

All radiosonde temperature profiles are interpolated to obtain a grid of values every 10 hPa of pressure (see Figure \ref{fig:DCM temperature pressure}). We then compute the Pearson correlation coefficient between the rate and temperature at each pressure level, obtaining around 100 values.

We chose to represent the temperature by pressure levels rather than by altitude, because the correlation between the rate and temperature is significantly stronger when expressed in pressure coordinates.

\begin{figure}
    \centering
    \includegraphics[width=0.66\linewidth]{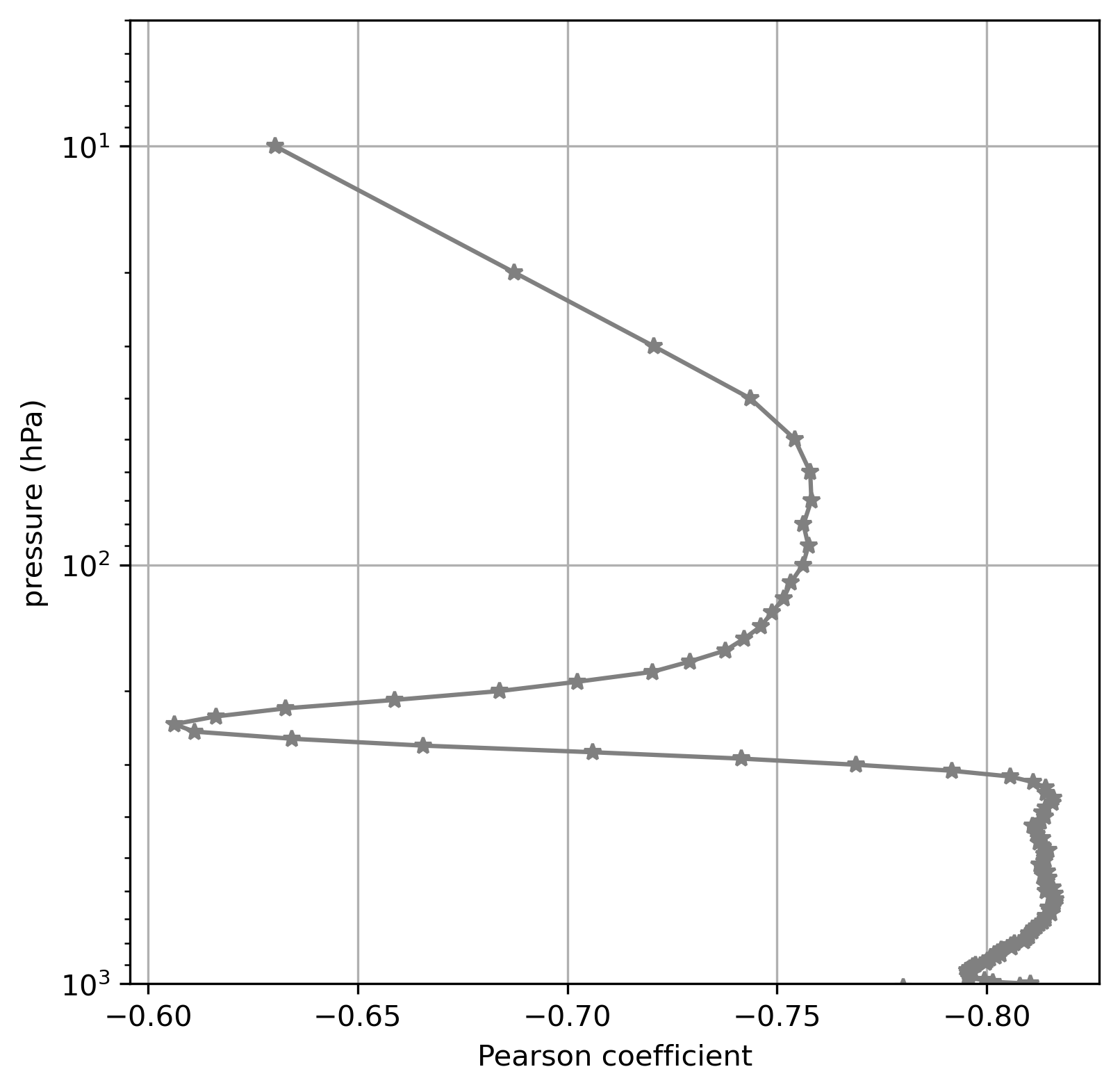}
    \caption{DCM method. Pearson coefficients of the correlation between the cosmic-muon rate and temperature for each pressure level versus the pressure level. Values above 10 hPa have been excluded, as the measurements are very sparse.}
    \label{fig:correlation pressure}
\end{figure}

The resulting correlation coefficients are shown in Figure \ref{fig:correlation pressure}. 

Observing the Pearson correlation curve, the difference from the weight calculated using the MSS method becomes immediately evident. In particular, the correlation shows a narrow minimum around 240 hPa, while the maximum correlation is found between 350 and 750 hPa. The origin of this structure is not yet fully understood, nor is it clear whether it is characteristic of this latitude or a more general feature.

\begin{figure} [hbt!]
    \centering
    \includegraphics[width=1\linewidth]{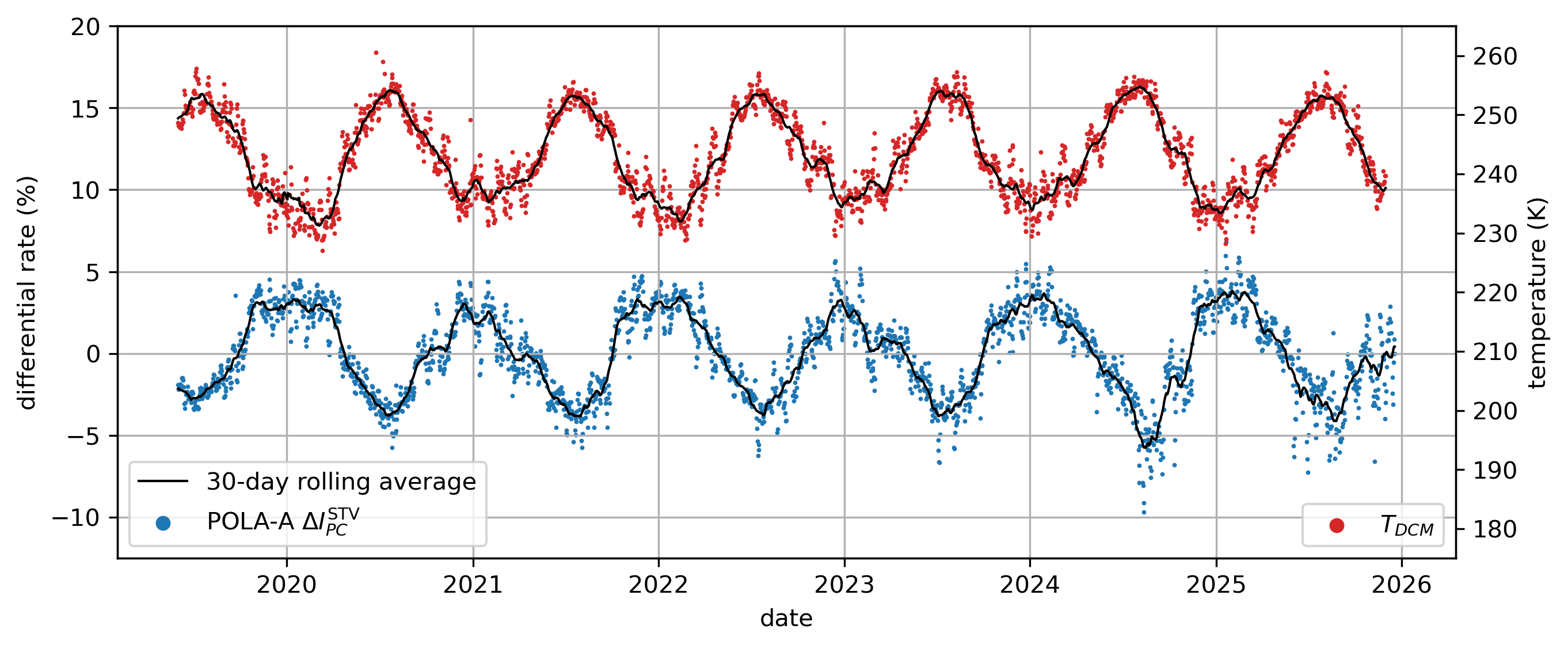}
    \caption{Temperature computed with method DCM (red). In blue, the differential rate for short-term variations. All 1-day time series are plotted together with a 30-day rolling average (black lines), shown to highlight the underlying trend.}
    \label{fig:tempDCM}
\end{figure}

With proper normalization, the function in Figure \ref{fig:correlation pressure} represents the specific weight function (as for equation \ref{eq:discrete}), from which the  temperature $T_{DCM}$ is derived, shown in Figure \ref{fig:tempDCM}. 

\begin{figure} [hbt!]
    \centering
    \includegraphics[width=1\linewidth]{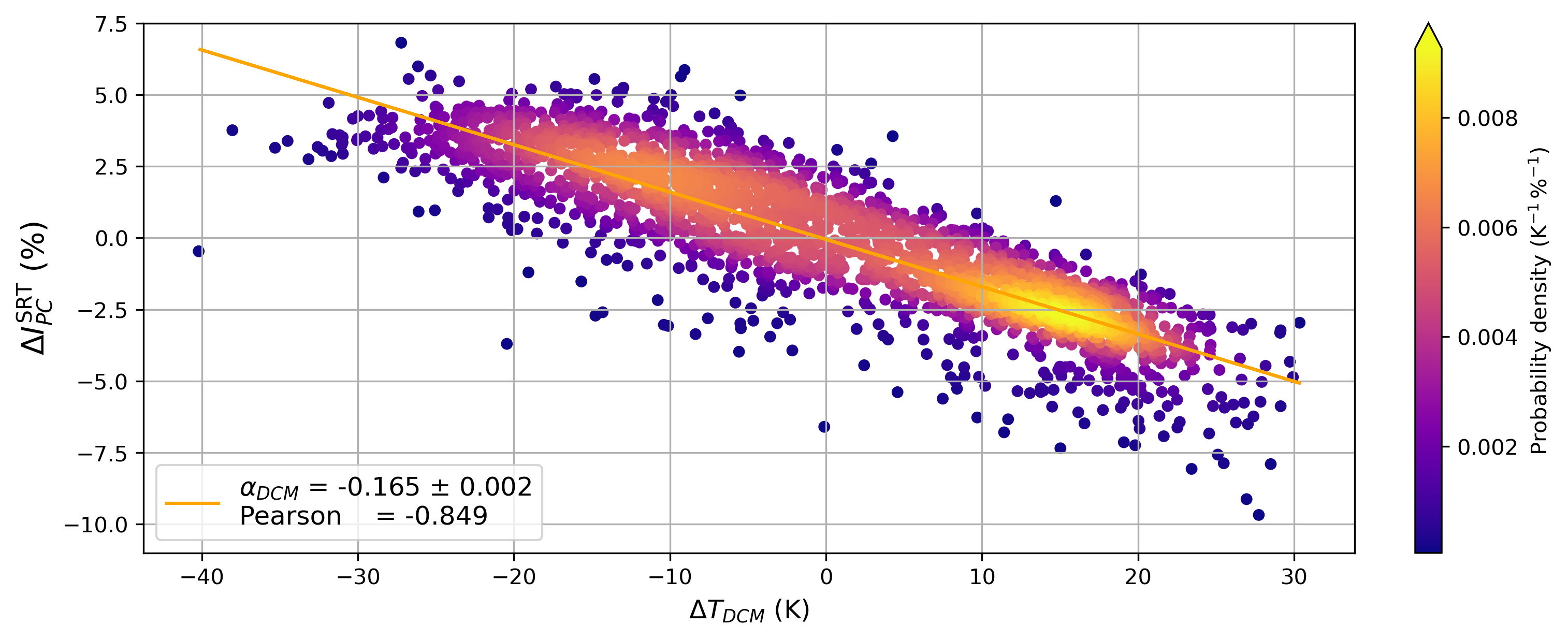}
    \caption{Correlation between $\Delta I_{PC}^{STV}$ and differential $T_{DCM}$.}
    \label{fig:pola_alpha_DCM}
\end{figure}

From this, we can calculate $\alpha_{DCM}$ according to the formula 
\begin{equation}
  \Delta I_{PC}^{STV} = \alpha_{DCM} * \Delta T_{DCM}
\end{equation}
where $\alpha_{DCM}$ units are $\%/K$.
The result of the linear fit of the correlation between the muon variation rate and the $\Delta T_{DCM}$ is reported in Figure \ref{fig:pola_alpha_DCM}.

\section{Comparison of temperature corrected rates  in Ny-Ålesund}

\begin{figure}
    \centering
    \includegraphics[width=1\linewidth]{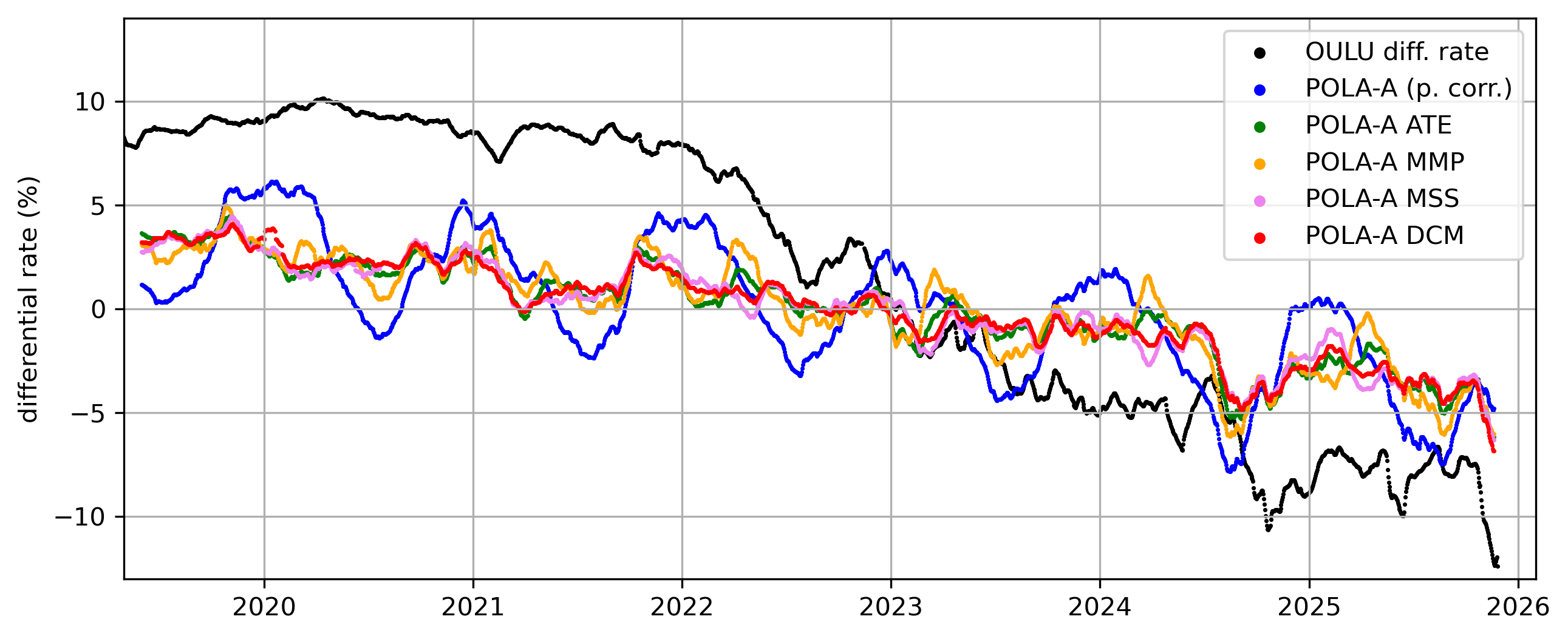}
    \caption{POLA-A differential rate corrected for pressure only (blue) together with the rates corrected for pressure and temperature using the four different methods described in this paper. The OULU differential neutron rate is also reported. }
    \label{fig:placeholder_for_comparison}
\end{figure}

The temperature coefficient values obtained for the four different models are reported in Table~\ref{tab:alpha}. The POLA-R differential rate, corrected for both pressure and temperature, is shown in Fig.~\ref{fig:placeholder_for_comparison} for the various methods considered, together with the original rate corrected for pressure only. All differential rates are expressed as percentages and represent deviations with respect to the five-year average over the entire observation period.

The different  methods are compatible within 1-1.5\% and the residual seasonal variation is considerably reduced. The neutron Oulu data, adjusted for both barometric pressure and efficiency,  are also reported and do not show any remaining seasonal variation. The decreasing trend due to the solar cycle is evident for both neutrons and muons, although the variation is much stronger in the neutron than in the POLA-R muon data.  

\begin{table}[h!]
\caption{POLA-A: Summary of the atmospheric temperature correction coefficients and Pearson correlation coefficients (between short-term variation differential rates and differential temperature or altitude), for the various models. } 
\label{tab:alpha}
\centering
\setlength{\tabcolsep}{20pt}
\begin {tabular}{| c | c | c | c |}
\toprule
$\alpha$ & Units & Result & Pearson coeff.\\
\midrule
 $\alpha_{ATE}$  & $\%/km$ & -4.475 $\pm$ 0.057 & -0.849  \\
\midrule
 $\alpha_{MMP}$  & $\%/K$ & -0.170 $\pm$ 0.003  & -0.756 \\
\midrule
 $\alpha_{MSS}$  & $\%/K$ & -0.165 $\pm$ 0.002 & -0.867 \\
\midrule
 $\alpha_{DCM}$  & $\%/K$ & -0.326 $\pm$ 0.003  & -0.891 \\
\bottomrule
\end{tabular}
\end{table}

\begin{figure}
    \centering
    \includegraphics[width=1\linewidth]{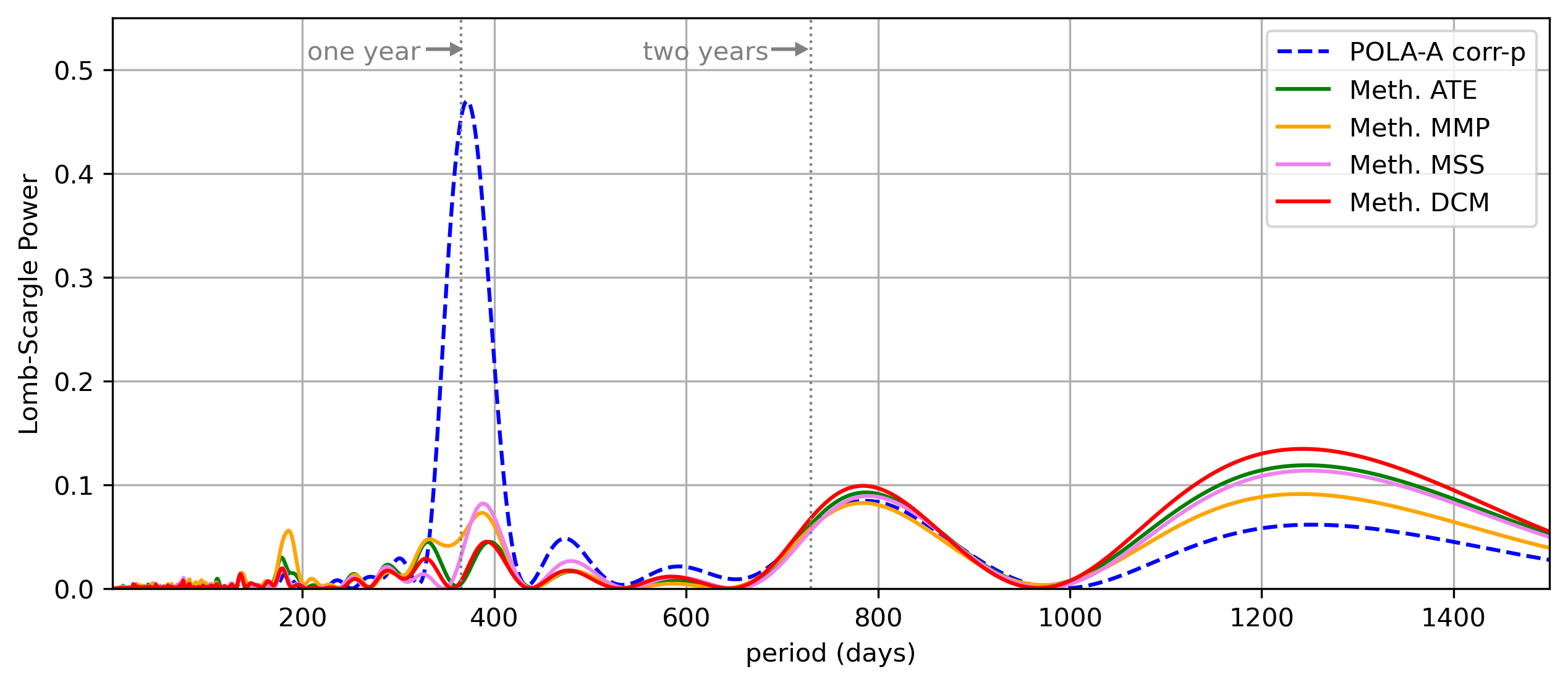}
    \caption{Periodograms of the POLA-R count rates corrected for pressure only (dashed curve) and for both pressure and temperature using the various methods described in this study. After applying the temperature correction, the prominent annual peak is suppressed, revealing spectral components at nearly two years and other longer-term periodicities.}
    \label{fig:placeholder_for_periodograms}
\end{figure}

To analyze the periodicities present in the muon rate time series in greater detail, we employ the Lomb–Scargle periodogram technique \cite{VanderPlas_2018}, following the approach adopted in our previous study \cite{POLAperiodicity}. This method is based on the optimization of sinusoidal components and is particularly well suited to time series with irregular sampling and data gaps.
After removing the temperature-dependent component of the atmospheric effect, the periodogram shows a substantial reduction of the annual periodicity (Figure  \ref{fig:placeholder_for_periodograms}), making it possible to investigate the presence of additional structures. However, a more detailed study of these features would require a larger dataset.

\section{Comparison with temperature coefficients at different locations}

The results for the temperature coefficient can be compared with those of the Global Muon Detector Network (GMDN) reported in \cite{deMendonça_2016} for four muon detectors located in Kuwait (KWT), Nagoya (NGY), Hobart (HBT), and S\~ao Martinho da Serra (SMS), based on data collected from January 2007 to December 2012. The two analyses have been performed in a similar way, but the POLA-R data are obtained at much higher latitude, thus in this way emphasizing the contribution of lower energies muons, which are better detected thanks to the reduced geomagnetic effect (see section \ref{sect:athmosphericeffects}).

\begin{figure}
    \centering
    \includegraphics[width=1\linewidth]{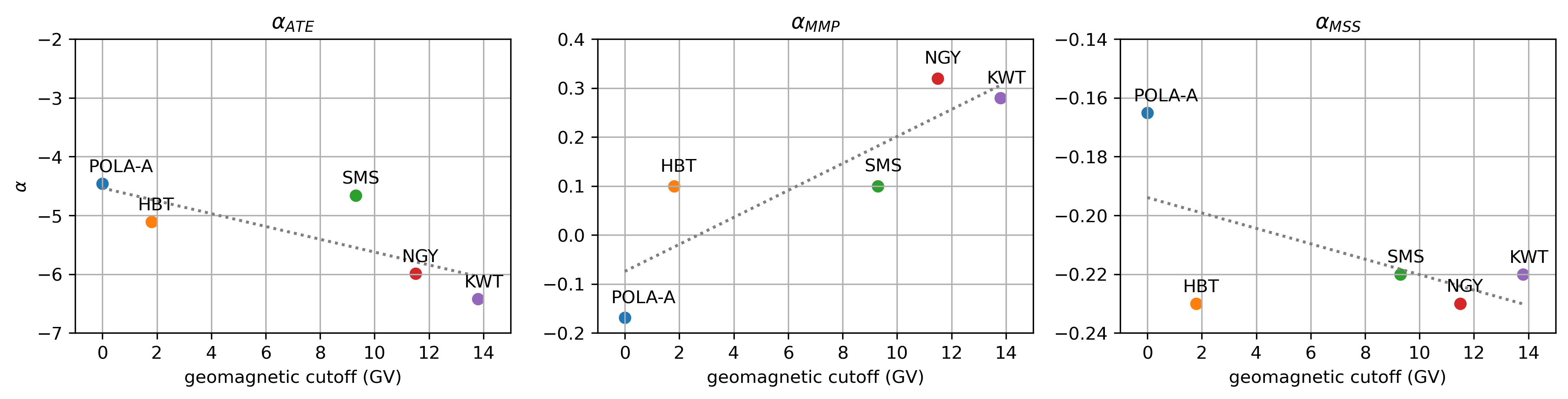}
    \caption{Comparison of the atmospheric temperature coefficients as a function of the geomagnetic cutoff in GMDN and POLA-R experiments.}
    \label{fig:comptempcoef}
\end{figure}

In Figure \ref{fig:comptempcoef} the measured coefficients are reported versus the geomagnetic cutoff. Within the uncertainties, the ATE and MMP results are qualitatively consistent with a linear trend. For the MSS method, the larger separation between the POLA-A and HBT data points leads to a less clear alignment with this trend.
It is worth to point out the different sign of the $\alpha_{{MMP}}$ coefficients between GMDN \cite{deMendonça_2016} and POLA-R,  probably indicating a greater sensitivity of the MMP method to different positive or negative correlation contributions at different latitudes. 

Overall, we believe that methods that use temperature-weighted contributions from multiple atmospheric layers are preferable to those relying on a single parameter, as they better account for the local atmospheric structure, which varies with latitude.

\section{Conclusions}

Three POLA-R scintillator detectors installed by the EEE Collaboration have been taking data for more than six years (from 2019 till 2025) at the Ny-Ålesund Research Station, on the Svalbard Islands (78.9° N latitude).

The data exhibit a pronounced seasonal variation, largely driven by atmospheric parameters. 
Such variations were corrected using daily radiosonde measurements collected by AWI directly above Ny-Ålesund. 

After applying different correction methods found in scientific literature, we propose an alternative technique  based on the  correlation between cosmic-muon rates and temperatures at different pressure levels. This method has the advantage of taking into account the local characteristics of the atmosphere and the latitude dependence of the atmospheric structure. Moreover, it also  proves capable to effectively correct the residual pressure dependence that the method based on surface pressure alone does not fully remove.

We also analyzed the residual periodic components of the muon rates and found that the dominant one-year oscillation is completely removed by the temperature correction, revealing smaller components that may be associated with different phenomena. Finally, we compared the corrected muon rates and the corresponding neutron  data showing a similar trend with time and solar cycle, although more marked in the neutron case. 

The temperature coefficients found with the different methods have been compared with available data at lower latitudes, qualitatively indicating a linear dependence of their magnitude with the geomagnetic cutoff.  

\section*{Acknowledgments}
The authors would like to thank the Alfred Wegener Institute for Polar and Marine Research (AWI) for providing the radiosonde data (RiS 1118).
We also would like to thank the Italian National Research Council (CNR) hosting the POLA-R detectors at their facilities in Ny-Ålesund, and the personnel of the Arctic Station Dirigibile Italia for their help in the management of the detectors.

We would also like to thank GMDN (Global Muon Detector Network) and the NMDB (Neutron Monitor DataBase) for providing access to the data.

\bibliography{EEE-bibliography}

\end{document}